\documentclass[12pt]{article}
\usepackage{fullpage}
\usepackage{amsmath}
\usepackage{amssymb}
\usepackage[dvips]{epsfig}

\def\##1{\underline{#1}}
\def\=#1{\underline{\underline{#1}}}

\def\+#1{\underline{\bf #1}}
\def\*#1{\underline{\underline{\bf #1}}}

\def\r#1{(\ref{#1})}
\def\l#1{\label{#1}}
\def\c#1{\cite{#1}}

\def\le{\left(}
\def\ri{\right)}
\def\les{\left[}
\def\ris{\right]}
\def\lec{\left\{}
\def\ric{\right\}}

\def\.{\mbox{ \tiny{$^\bullet$} }}

\def\epso{\epsilon_{\scriptscriptstyle 0}}

\def\muo{\mu_{\scriptscriptstyle 0}}

\def\ko{k_{\scriptscriptstyle 0}}
\def\co{c_{\scriptscriptstyle 0}}
\def\Eo{\#E_{\, \scriptscriptstyle 0}}
\def\Ho{\#H_{\, \scriptscriptstyle 0}}

\def\eps{\epsilon}

\def\ct{\cos\theta}
\def\st{\sin\theta}

\def\ux{\hat{\#u}_x}
\def\uy{\hat{\#u}_y}
\def\uz{\hat{\#u}_z}

\begin{document}

\begin{center}

{\bf {\Large  Towards gravitationally assisted negative refraction of light  by vacuum
}}

 \vspace{10mm} \large

 Akhlesh  Lakhtakia\footnote{Fax: +1 814 863 4319; e--mail: akhlesh@psu.edu; also
 affiliated with Department of Physics, Imperial College, London SW7 2 BZ, UK}\\
 {\em CATMAS~---~Computational \& Theoretical
Materials Sciences Group\\ Department of Engineering Science and
Mechanics\\ Pennsylvania State University, University Park, PA
16802--6812, USA}\\
\bigskip
Tom G. Mackay\footnote{Corresponding Author. Fax: + 44 131
650 6553; e--mail: T.Mackay@ed.ac.uk.}\\
{\em School of Mathematics,
University of Edinburgh, Edinburgh EH9 3JZ, UK}\\

\end{center}

\vspace{4mm}

\normalsize

\begin{abstract}

Propagation of electromagnetic plane waves in some directions in gravitationally affected vacuum over limited ranges of
spacetime can be such that the phase velocity vector casts a negative projection on the
time--averaged Poynting vector. This conclusion suggests, {\em inter alia\/}, gravitationally assisted negative refraction by vacuum.

\end{abstract}


\section{Introduction}
The discovery of (purportedly) isotropic, homogeneous, dielectric--magnetic materials that bend 
electromagnetic rays the ``wrong way"  \c{wrongway} created quite a stir in 2001 \c{SSS},
with claims and counterclaims flying all around \c{Garcia,Valanju}. The situation  has recently 
been settled,
with unequivocal demonstrations by several independent groups \c{NPV_expt1}--\c{PS04}. See Ref. 8
for a comprehensive
review.
A range of exotic and potentially useful phenomenons~---~such as
negative
refraction, negative Doppler shift and  inverse \u{C}erenkov
radiation~---~have been predicted for materials of this type, wherein the phase velocity
is directed in opposition to the energy velocity as quantified through the time-averaged
Poynting vector. These materials have several names, including left--handed materials,
negative--index materials, and negative-phase--velocity (NPV) materials. We prefer the last term \c{LMW}.

Subsequently, the possibility of NPV propagation of light
and other electromagnetic waves was established in a variety of anisotropic materials \c{HC02}--\c{ML_PRE}.
In these materials, NPV propagation is indicated by the projection of the phase velocity on the time--averaged
Poynting vector being negative.

Even more interestingly, materials that do not permit the observation of NPV propagation by    observers 
in a relatively stationary (i.e., co--moving) inertial reference frame have been shown, after the
invocation of the postulates of special theory of relativity (STR), to allow observation
of NPV propagation in other
inertial frames \c{ML04a}. That permits one to
envisage STR negative
refraction being exploited  in astronomical scenarios such as, for example, in
the remote sensing of planetary and asteroidal surfaces from space stations.
Quite possibly, space telemetry technologies will be the first to reap the
benefits of STR negative refraction. Application to remotely guided, extraterrestrial mining and 
manufacturing industries can also be envisioned. Furthermore, many
unusual astronomical phenomenons would be discovered and/or explained
via  STR negative refraction to 
interpret data
collected via telescopes \c{ML04b}.

As is well known, vacuum (i.e., matter--free space) appears the same to all inertial observers \c{Chen}. Therefore,
as a co--moving observer cannot deduce the occurrence of NPV propagation in vacuum, neither can any
observer moving with a constant velocity. This could lead one to believe that NPV propagation is impossible
in huge expanses of interstellar space. However, gravitational fields from nearby massive objects will
certainly distort electromagnetic propagation, which is a principal consequence of the general theory of
relativity and is indeed used nowadays in GPS systems. Our objective here is to establish that gravitationally affected
vacuum can support NPV propagation, at least in spacetime manifolds of limited extent.

\section{Theory}

A gravitational field curves spacetime, which effect is captured through a metric $g_{\alpha\beta}$.\footnote{Roman 
indexes take the values 1, 2 and 3; while Greek indexes take the
values 0, 1, 2, and 3.}
Electromagnetic
propagation in gravitationally affected vacuum may be described in terms of propagation in an instantaneously 
responding medium in flat spacetime \c{Skrotskii,Plebanski}, 
at least in spacetime manifolds of limited extent. That is,
the {\em nonuniform\/} metric $g_{\alpha\beta}$ can be locally approximated 
by the {\em uniform\/} metric $\tilde{g}_{\alpha\beta}$ \c{FN}.
On asssuming the convention  $\tilde{g}_{\alpha \beta} = (+,-,-,-)$,
the constitutive relations of vacuum in the equivalent flat
spacetime are expressed in Gaussian units as \c{Plebanski}
\begin{eqnarray}
D_\ell &=& \eps_{\ell m} E_m + \eps_{\ell mn} \tilde{g}_m H_n \,, \l{Di}\\
B_\ell &=&    \mu_{\ell m} H_m - \eps_{\ell mn} \tilde{g}_m  E_n\,, \l{Bi}
\end{eqnarray}
where $\eps_{\ell mn}$ is the Levi--Civita tensor, and
\begin{eqnarray}
\eps_{\ell m} &=& \mu_{\ell m}
= - \le -\tilde{g} \ri^{1/2} \, \frac{\tilde{g}^{\ell m}}{\tilde{g}_{00}}\,,\\
\tilde{g}_\ell &=&  \frac{\tilde{g}_{0\ell}}{\tilde{g}_{00}}\,,
\end{eqnarray}
with $\tilde{g} = \mbox{det} \les \,\tilde{g}_{\alpha \beta} \, \ris$.  We note that the metric 
$\tilde{g}_{\alpha \beta}$
is real symmetric \c{Sachs}.

 The constitutive relations \r{Di} and \r{Bi}
can be expressed in  3--vector form as
\begin{eqnarray}
\#D &=& \epso \=\gamma \. \#E -  \#\Gamma \times \#H
\,, \l{D_cr} \\
\#B &=& \muo \=\gamma \. \#H + \#\Gamma  \times \#E
\,, \l{B_cr}
\end{eqnarray}
wherein SI units are implemented. The scalar constants $\epso$ and
$\muo$ denote the permittivity and permeability of  vacuum in the absence of a gravitational field, 
respectively, and $\co = \sqrt{1/ \epso \muo }$. Our coordinate
system is chosen such that the second--rank cartesian tensor $\=\gamma$ is diagonal; i.e., $\=\gamma
= \mbox{diag} \, \le \gamma_x, \gamma_y, \gamma_z \ri$. In the
gravitational field of a mass rotating with angular momentum
$\#J$, the gyrotropic vector $\#\Gamma $ is proportional to $\#R
\times \#J$, where $\#R$ is the radial vector from the centre of
mass to the point of observation. By considering a small region of
space at a sufficiently remote location from the centre of mass,
we take  $\#\Gamma$ to be independent of $\#R$.

We seek planewave solutions
\begin{eqnarray}
\#E &=& {\rm Re} \lec\Eo \exp \les i \le  \#k \. \#r - \omega t \ri \ris\ric\,,
\\   \#H &=& {\rm Re} \lec\Ho \exp \les i \le  \#k \. \#r - \omega t \ri
\ris\ric\,,
  \l{pw}
\end{eqnarray}
to the source--free Maxwell curl postulates
\begin{eqnarray}
&& \nabla \times \#E + \frac{\partial}{\partial t} \#B = \#0\,,\\
&& \nabla \times \#H - \frac{\partial}{\partial t} \#D = \#0\,.
\end{eqnarray}
Here $\#k$ is the wavevector, $\#r$ is the position vector,
$\omega$ is the angular frequency, and $t$ denotes the time;
whereas $\Eo$ and $\Ho$ are complex--valued amplitudes.

An eigenvector equation for $\Eo$ is developed as follows. By
combining \r{D_cr}--\r{pw} with the Maxwell curl postulates, we
derive
\begin{eqnarray}
\#p \times \Eo &=& \omega \muo \=\gamma \. \Ho \,, \l{pE0} \\
\#p \times \Ho &=& - \omega \epso \=\gamma \.  \Eo \,, \l{pH0}
\end{eqnarray}
in terms of
\begin{equation}
\#p = \#k - \omega \#\Gamma \,. \l{p}
\end{equation}
The use of \r{pE0} to  eliminate $\Ho$ from \r{pH0} provides, after
some simplification,
\begin{equation}
\=W \. \Eo = \#0 \,, \l{WE}
\end{equation}
where
\begin{equation}
\=W =  \le \, \ko^2 \mbox{det} \, \les \, \=\gamma \, \ris  - \#p
\. \=\gamma \. \#p \, \ri \=I + \#p \, \#p \. \=\gamma\,,
\end{equation}
and the notation $\ko = \omega \sqrt{\epso \muo}\,$
has been introduced.  A dispersion relation thus emerges from
\r{WE} as
\begin{equation}
\mbox{det} \, \les \, \=W \, \ris = 0\,,
\end{equation}
which may be expressed in the form
\begin{equation}
\ko^2 \,  \mbox{det} \, \les \, \=\gamma \, \ris \le \ko^2 \,
\mbox{det} \, \les \, \=\gamma \, \ris - \#p\.\=\gamma\.\#p  \ri^2
= 0\,. \l{disp2}
\end{equation}
Hence, we conclude that planewave solutions  satisfy the condition
\begin{equation}
\#p\.\=\gamma\.\#p = \ko^2 \, \mbox{det} \, \les \, \=\gamma \,
\ris\,. \l{disp_cond}
\end{equation}

Let us consider eigenvector solutions to \r{WE}. Substitution of
\r{disp_cond} into \r{WE} provides
\begin{equation}
\#p \, \#p \. \=\gamma \. \Eo = \#0\,;
\end{equation}
thereby, all eigenvector solutions are necessarily  orthogonal to
$\#p\.\=\gamma$. To proceed further, let us~---~without any loss
of generality~---~choose the wavevector $\#k$ to lie
along  $z$ axis, and the   vector $\#\Gamma$ to lie in the $xz$
plane; i.e.,
\begin{eqnarray}
\#k &=& k \uz\,, \\
\#\Gamma &=& \Gamma \le \ux \st +\uz\ct \ri\,,
\end{eqnarray}
where $\ux$, $\uy$ and $\uz$ are the cartesian unit vectors in the
equivalent flat spacetime.
Since
\begin{eqnarray}
\#p \. \=\gamma &=&  - \omega \Gamma \gamma_x \st \ux + \le k
 - \omega \Gamma  \ct \ri \gamma_z\uz\,,
\end{eqnarray}
it is clear that  two linearly independent eigenvectors
satisfying \r{disp_cond} may be stated as
\begin{eqnarray}
\#e_{\,1} &=& \uy \,, \l{e1} \\
\#e_{\,2} &=& \uy  \times \le \#p \. \=\gamma \ri  \\
&=& \le k  - \omega \Gamma  \ct \ri \gamma_z \ux + \omega
\Gamma \gamma_x \st \uz \,. \l{e2}
\end{eqnarray}
After assuming that $\=\gamma$ is invertible, we deduce
the corresponding magnetic field eigenvectors  from
\r{pE0} as
\begin{eqnarray}
\#h_{\,1} &=& \frac{1}{\omega \muo}
 \=\gamma^{-1} \. \les \le \omega \Gamma \ct - k \ri \ux - \omega
 \Gamma \st \uz \,\ris
 \,, \l{h1} \\
\#h_{\,2} &=& \frac{1}{ \omega \muo } \=\gamma^{-1} \. \les \le k
- \omega \Gamma \ct \ri^2 \gamma_z + \le \omega \Gamma \st \ri^2
\gamma_x \ris \, \uy
 \,. \l{h2}
\end{eqnarray}
Hence, the general solution is given by
\begin{eqnarray}
\Eo &=& C_1 \uy + C_2 \les  \le k  - \omega \Gamma
  \ct \ri \gamma_z \ux + \omega \Gamma \gamma_x \st \uz \ris
\,, \l{E_vec} \\
\Ho &=& \frac{1}{\omega \muo}
 \=\gamma^{-1} \.  \Bigg\{ C_1 \,
 \les \le \omega \Gamma \ct - k \ri \ux - \omega
 \Gamma \st \uz \,\ris  \nonumber \\ && + C_2
 \les \le k - \omega \Gamma \ct
\ri^2 \gamma_z + \le \omega \Gamma \st \ri^2 \gamma_x \ris \, \uy
\, \Bigg\}
 \l{H_vec} \,,
\end{eqnarray}
wherein $C_1$ and $C_2$ are arbitrary constants.

 The wavenumbers arise  from  the dispersion relation \r{disp2} as follows.
Substituting \r{p} into \r{disp_cond}, we obtain the $k$--quadratic
expression
\begin{equation}
k^2 \gamma_z  - 2 k \gamma_z \omega \Gamma  \ct + \omega^2
\Gamma^2 \le \gamma_x \sin^2 \theta + \gamma_z \cos^2 \theta \ri
 - \ko^2\,  \mbox{det} \, \les \,\=\gamma \,\ris  = 0\,, \l{quad}
\end{equation}
since $\#k \. \#\Gamma = k \Gamma \,\ct $. The two $k$--roots of
\r{quad} are
\begin{eqnarray} k^+ &=& \omega \l{kp_root}
 \le \Gamma \ct + \sqrt{ \epso \muo \gamma_x \gamma_y
 -\frac{\gamma_x}{\gamma_z} \Gamma^2 \sin^2 \theta  } \ri\,,\\
k^- &=& \omega \l{km_root}
 \le \Gamma \ct - \sqrt{ \epso \muo \gamma_x \gamma_y
 -\frac{\gamma_x}{\gamma_z} \Gamma^2 \sin^2 \theta  } \ri\,.
 \end{eqnarray}

Finally, let us consider the time--averaged Poynting vector given by
\begin{eqnarray}
\#P &=& \frac{1}{2}\, {\rm Re}\lec \Eo\times \Ho^\ast\ric\,.
\l{P_def}
\end{eqnarray}
After utilizing the general solution \r{E_vec} and \r{H_vec}, the
component of the Poynting vector aligned with the $\uz$ axis is
obtained as
\begin{eqnarray}
\hat{\#u}_z \. \#P &=&  \frac{1}{2\,\omega \muo \gamma_z  } \le k
- \omega \Gamma \ct \ri \le |C_1|^2 + |C_2 |^2 \gamma_z \omega^2
\mbox{det}\,\les \, \=\gamma \, \ris \ri .
\end{eqnarray}

The energy density flow in the direction of the wavevector
$\#k^+$, corresponding to the root $k^+$ given in \r{kp_root}, is thus
\begin{eqnarray}
\#k^+ \. \#P &=&  \frac{1}{2 \muo \gamma_z } \les \Gamma \ct
\sqrt{ \epso \muo \gamma_x \gamma_y
 -\frac{\gamma_x}{\gamma_z} \Gamma^2 \sin^2 \theta  } + \le
 \epso \muo \gamma_x \gamma_y
 -\frac{\gamma_x}{\gamma_z} \Gamma^2 \sin^2 \theta
  \ri \ris \nonumber \\ && \times \le |C_1|^2 + |C_2 |^2 \gamma_z \omega^2
\mbox{det}\,\les \, \=\gamma  \, \ris \ri \,.
\end{eqnarray}
Let us notice that the inequality
\begin{equation}
\epso \muo \gamma_x \gamma_y
 -\frac{\gamma_x}{\gamma_z} \Gamma^2 \sin^2 \theta  \ge  0
 \end{equation}
 must be fulfilled in order for $k$ to be real--valued. Therefore, the
 defining inequality for NPV propagation, namely
 \begin{equation}
\#k^+ \. \#P < 0\,,
\end{equation}
is satisfied provided that
\begin{equation}
- \Gamma \ct > \sqrt{ \epso \muo \gamma_x \gamma_y
 -\frac{\gamma_x}{\gamma_z} \Gamma^2 \sin^2 \theta  }
 \label{NPVp}
 \end{equation}
 holds.
Analogously,  we find that NPV propagation is signalled  for the
$k^-$ wavenumber  by the condition
\begin{equation}
 \Gamma \ct > \sqrt{ \epso \muo \gamma_x \gamma_y
 -\frac{\gamma_x}{\gamma_z} \Gamma^2 \sin^2 \theta  }.
 \label{NPVn}
 \end{equation}
 In deriving \r{NPVp} and \r{NPVn}, we used the fact that $\gamma_{x,y,z} > 0$ by virtue of the
 signature of $\tilde{g}_{\alpha\beta}$.

We note that the NPV conditions \r{NPVp} and \r{NPVn} are independent of frequency.

\section{Concluding Remarks}
The inequalities \r{NPVp} and \r{NPVn} can be satisfied for specific
ranges of the angle $\theta$, for given $\=\gamma$ and $\Gamma$.
Thus, we have shown that NPV propagation in some directions is
possible in gravitationally affected vacuum over limited ranges of
spacetime. The possible existence of gravitational fields which can deliver $\=\gamma$ and $\Gamma$
necessary for the satisfaction of  \r{NPVp} and/or \r{NPVn}
is a matter for astrophysicists to discuss. 

We are content here to state that, just as scientific
and technological applications of STR negative refraction (by materials) can be envisaged \c{ML04a,ML04b},
similar and different consequences of gravitationally 
assisted negative refraction by vacuum are possible.
In particular, designers of channels for space communication shall also have to account for the
possibility of negative refraction due to massive objects between the two ends of every channel.

Furthermore,
current ideas on the distribution of mass in the as--observed universe may require
significant revision, since electromagnetic signals from distant objects
may be deflected in a manner which has not hitherto been accounted
for.  Thus,
our work has implications for gravitational
lenses \c{KB04}. Gravitational lensing involves nonuniform metrics,
and the distribution of matter in the
universe has been constantly changing.
While
the
spatiotemporally \emph{local} evolution
of the universe may be deduced  adequately from electromagnetic signals
received by our telescopes, reasonably accurate
deductions
about the spatiotemporally \emph{global} evolution of
the universe from similar measurements may be particularly difficult to make~---~owing
 to gravitationally assisted negative refraction.
 
Finally, our work suggests that it is time that research on consequences
of gravitationally assisted  negative refraction by materials be undertaken.

\medskip

\noindent{\bf Acknowledgement.} We gratefully acknowledge discussions with Dr. Sandi Setiawan 
(University of Edinburgh),  Dr. Chandra  Roychoudhuri (University
of Connecticut) and Mr. James McIlroy (Imperial College). TGM thanks the Nuffield Foundation for supporting
 his visit to Pennsylvania State University.

\end{document}